\documentclass{ws-ijmpa}
\usepackage{cite}
\usepackage{xcolor}
\usepackage[verbose,hypertexnames=false]{hyperref}
\hypersetup{colorlinks=false,allbordercolors=blue,pdfborderstyle={/S/U/W 1}}

\usepackage{mathrsfs}
\usepackage{amsmath}
\usepackage{amssymb}
\usepackage{braket}
\usepackage{graphicx}
\usepackage{xcolor}
\usepackage{bm}
\usepackage{comment}

\allowdisplaybreaks[4]

\begin{document}
\markboth{Naoki Fukushima and Kei-Ichi Kondo}{Restoration of residual gauge symmetries towards color confinement due to topological defects}

\title{Restoration of residual gauge symmetries \\
 towards color confinement due to topological defects} 

 \author{Naoki Fukushima}

 \address{Department of Physics, Graduate School of Science and Engineering, Chiba University, Chiba 263-8522, Japan\\
 fnuakouksih9i1m3a@chiba-u.jp}

 \author{Kei-Ichi Kondo}

 \address{Department of Physics, Graduate School of Science, Chiba University, Chiba 263-8522, Japan\\
 kondok@faculty.chiba-u.jp}


\date{\today}
\maketitle

\begin{abstract}
  We reconsider the restoration of the residual gauge symmetry (RGS) due to topological effects as a possible criterion for color confinement. Although the RGS is ``spontaneously broken'' in the perturbative vacuum, it must be restored in the true confining vacuum of QCD, provided that color confinement phase is a disordered phase where all of symmetries are unbroken. Therefore, the disappearance of the massless Nambu-Goldstone pole associated with this spontaneously breaking can be regarded as a criterion for color confinement.
  In the Lorenz gauge, indeed, the restoration condition was shown to agree with the Kugo-Ojima color confinement criterion at least for a special choice of the residual gauge transformation.
  In the previous paper, we have proposed to generalize this idea by including the topological defects.
  In this paper, we elaborate this scenario and obtain the criterion (i) by examining a finite large gauge transformation to properly take into account the topological effects and to specify the RGS, and (ii) by obtaining the condition for restricting the possible topological configurations so that they give a finite Euclidean action to give a non-vanishing contribution to the path integral.
\end{abstract}


\section{Introduction}

Color confinement is still an unsolved important problem in particle physics. Especially, quark confinement is well understood based on the dual superconducting picture where condensation of magnetic monopoles and anti-monopoles occurs\cite{dual-superconductor}.
On the other hand, gluon confinement is not so well understood in this picture.
However, a general color confinement criterion exists:

According to the analysis of Kugo and Ojima \cite{KO} based on the manifestly Lorentz covariant canonical operator formalism, no particles with color charges can be observed, if the Kugo-Ojima (KO) criterion\cite{KO} is satisfied for the particular value $u(0)$ of the infrared limit $k \to 0$ of a specific function called the KO function $u(k)$ \cite{KO-prove}.

The local gauge symmetry remaining even after imposing a gauge fixing condition is called the \textit{residual gauge symmetry} (RGS). Although the RGS is ``spontaneously'' broken in the perturbative vacuum, it is expected to be restored in the true confining vacuum.
The criterion for the restoration can be formulated as the condition for disappearance of the massless Nambu-Goldstone (NG) pole associated with the ``spontaneous symmetry breaking'' of the RGS.
Remarkably, the KO criterion was reproduced as a restoration condition for the RGS in the Lorenz gauge when the gauge transformation function is a linear function of the spacetime coordinate $x^\mu$, as shown by Hata\cite{Hata-method}.
Here it should be remarked that the KO criterion was derived only in the Lorenz gauge and is clearly gauge-dependent.
Therefore, it cannot be directly applied to the other gauge-fixing conditions.

The restoration condition for the RGS can also be extended to more general gauge transformations involving topological configurations.
In the previous paper\cite{Kondo-Fukushima}, indeed, we introduced topological defects to demonstrate in the configuration space that it is possible to satisfy the restoration condition even in the Maximal Abelian gauge.

Hence, it enables us to understand color confinement from the view point of restoration of the RGS due to topological defects also for the gauges other than the Landau gauge.
However,
the infinitesimal gauge transformation adopted in the previous paper is not sufficient to correctly take into account topological configurations as the RGS.
Moreover, some topological configurations introduced in the previous paper give an infinite Euclidean action without any regularizations and do not contribute to the path integral in the naive way.

  In this paper, we elaborate our scenario to obtain the criterion in the following way:
\begin{enumerate}
\item[(i)] We use a finite gauge transformation to properly take into account the topological effects and to specify the RGS.

\item[(ii)] We restrict the possible topological configurations so that they give a finite Euclidean action to give a non-vanishing contribution to the path integral.
\end{enumerate}


This paper is organized as follows.
In Sec. \ref{sec:KO}, we give a review on the Kugo-Ojima color confinement criterion  in the Lorenz gauge from the viewpoint of the restoration of the RGS  supplemented with the current status of the related investigations.
In Sec. \ref{sec:residual-Lorenz}, we extend the restoration criterion of the RGS in the Lorenz gauge by including the topological configurations resulting from a  finite gauge transformation.
In Sec \ref{sec:topological-configuration}, we show that there exists the RGS related to topological configuration obtained by a finite gauge transformation with the help of the Witten Ansatz and that the relevant topological configurations give a finite action integral to contribute to the path integral.
The final section is devoted to conclusion and discussion.
The details of the calculations are given in Appendix \ref{sec:EOM}.

\section{\label{sec:KO} Kugo-Ojima color confinement criterion and  restoration of the RGS in the Lorenz gauge}

In this section we review the relationship between the Kugo-Ojima color confinement criterion and the restoration of the RGS in the Lorenz gauge, which is supplemented with the current status for the verification of the criterion in the numerical simulations to explain the motivation of this paper.

We consider the Yang-Mills theory coupled to the matter field
where the Yang-Mills gauge field $\mathscr{A}_\mu(x)$ is defined as the Lie algebra valued field $\mathscr{A}_\mu(x)=\mathscr{A}_\mu^A(x)T_A$ with the generators $T_A$ ($A=1, \cdots ,  \text{dim} SU (N) =N^2 - 1$) of the Lie algebra $su(N)$ for the $SU (N)$ gauge group.
The total Lagrangian density is given by
\begin{align}
  \mathscr{L} = \mathscr{L}_{\text{inv}} + \mathscr{L}_{\text{GF+FP}}.
  \label{eq:total-lagrangian}
\end{align}
The first term $\mathscr{L}_{\text{inv}}$ is the gauge-invariant part given by
\begin{align}
  \mathscr{L}_{\text{inv}} = - \frac{1}{4} \mathscr{F}_{\mu \nu} \cdot \mathscr{F}^{\mu \nu} + \mathscr{L}_{\text{matter}} (\varphi , D_\mu \varphi),
\end{align}
with the Lie algebra valued field strength $\mathscr{F}_{\mu \nu}$ of the gauge field $\mathscr{A}_\mu$ defined by
$\mathscr{F}_{\mu \nu} := \partial_\mu \mathscr{A}_\nu - \partial_\nu \mathscr{A}_\mu - i g [\mathscr{A}_\mu , \mathscr{A}_\nu] = - \mathscr{F}_{\nu \mu}$
 and the covariant derivative $D_\mu \varphi$ defined by $D_\mu \varphi := \partial_\mu \varphi - i g \mathscr{A}_\mu \varphi$
 for the matter field $\varphi$ in the fundamental representation.

 The second term $\mathscr{L}_{\text{GF+FP}}$ is the sum of the gauge fixing (GF) term $\mathscr{L}_{\text{GF}}$ including the Nakanishi-Lautrup (NL) field $\mathscr{B} (x)$ as the Lagrange multiplier field to incorporate the gauge fixing condition to be specified later, and the associated Faddeev-Popov (FP) ghost term $\mathscr{L}_{\text{FP}}$ including the ghost field $\mathscr{C} (x)$ and the antighost field $\bar{\mathscr{C}} (x)$. Here $\mathscr{B} (x),  \mathscr{C} (x), \bar{\mathscr{C}} (x)$ are also Lie algebra valued fields.

First, we consider the \textit{Lorenz gauge fixing} given by the usual local condition:
 \begin{align}
   \partial^\mu \mathscr{A}_\mu^A(x) = 0 .
 \end{align}
Even after imposing the gauge fixing condition which breaks the gauge symmetry, the gauge theory with the total Lagrangian density $\mathscr{L}$ has the Becchi-Rouet-Store-Tyutin (BRST) symmetry.
Using the nilpotent BRST transformation $\delta_{\text{B}}$, indeed, the associated GF+FP term for the Lorenz gauge is written in the BRST exact form:
\begin{align}
  \mathscr{L}_{\text{GF+FP}} = - i \delta_{\text{B}} \left[\text{tr} \left\{\bar{\mathscr{C}} (\partial^\mu \mathscr{A}_\mu + \frac{\alpha}{2} \mathscr{B})\right\}\right]
\end{align}
where $\alpha$ is the gauge-fixing parameter.
Therefore, $\delta_{\text{B}} \mathscr{L}_{\text{GF+FP}} = 0$ due to nilpotency $\delta_{\text{B}}^2 \equiv 0$.

 Let $\mathcal{V}_{\text{phys}}$ be the physical subspace of the total state space $\mathcal{V}$ with an indefinite metric defined using the nilpotent BRST charge $Q_B$ as the generator of the BRST transformation:
\begin{align}
  \mathcal{V}_{\text{phys}} = \{\ket{\text{phys}} \in \mathcal{V} ; \ Q_B \ket{\text{phys}} = 0\} \subset \mathcal{V}.
\end{align}
 Kugo and Ojima \cite{KO} claimed: if the following condition called the \textit{Kugo-Ojima color confinement criterion} in the Lorenz gauge is satisfied:
\begin{align}
  \lim_{p^2 \rightarrow \ 0} u^{A B} (p^2) = - \delta^{A B}
 \ \Leftrightarrow \
  0 = \lim_{p^2 \rightarrow \ 0} [\delta^{A B} + u^{A B} (p^2)] ,
  \label{eq:KO-criterion}
\end{align}
for the function $u^{A B}$ called the \textit{Kugo-Ojima function} defined by
 \begin{align}
   u^{A B} (p^2) \left(g_{\mu \nu} - \frac{p_\mu p_\nu}{p^2}\right) = \int d^D x \ e^{i p (x - y)} \braket{0 | \text{T} [(\mathscr{D}_\mu \mathscr{C})^A (x) g (\mathscr{A}_\mu \times \bar{\mathscr{C}})^B (y)] | 0},
 \end{align}
 then the color charge operator $Q^A$ is well defined, namely, the color symmetry is not spontaneously broken and $Q^A$ vanishes for any physical state $\Phi , \Psi \in \mathcal{V}_{\text{phys}}$:
 \begin{align}
   \braket{\Phi | Q^A | \Psi} = 0 \ \text{for} \ \Phi , \Psi \in \mathcal{V}_{\text{phys}}.
 \end{align}
In what follows we introduce the notation: $(\mathscr{A} \times \mathscr{B})^A = f^{ABC} \mathscr{A}^B \mathscr{B}^C$  for the Lie-algebra valued quantities $\mathscr{A}, \mathscr{B}$ using the structure constant $f^{ABC}$ for the $su (N)$ Lie algebra with the indices $A, B, C = 1, \cdots , \text{dim} SU (N) = N^2 - 1$.
The Kugo-Ojima result means that all colored particles cannot be observed and only color singlet particles can be observed, because the BRS singlets as physical particles are all color singlets, while colored particles belong to the BRS quartet representation  and cannot be observed due to zero-norm combinations.

In order to check whether the Kugo-Ojima criterion \eqref{eq:KO-criterion} really holds or not, the numerical simulations have been performed  on the lattice \cite{Furui-Nakajima,Sternbeck,Aguilar} where the Kugo-Ojima function $u(p^2)$ ($u^{AB}(p^2)=\delta^{AB}u(p^2)$) was measured in the Euclidean region.
In order to check the KO criterion, we need the value of the KO function in the infrared limit $p^2 \rightarrow \ 0$.
Taking the limit $p^2 \rightarrow \ 0$ needs the infinite volume lattice, which is impossible  to be realized in practice.
Therefore, the data $u(p^2)$ at small but non-zero $p^2$ obtained on the large but finite lattice were extrapolated to $p^2 \rightarrow 0$.
In those works, it has been shown that the KO criterion is not exactly satisfied as a result of the extrapolation, although the Kugo-Ojima function $u(p^2)$ tends to approach  $- 1$ as the volume of the lattice increases.

However, the simple comparison between the theoretical result for the KO criterion in the continuum non-compact formulation and the numerical result of  simulations based on the lattice gauge theory is meaningless even after putting aside the extrapolation ambiguity and the finite volume effect.
The examination must be done with great care as explained below.

(i) incomplete gauge fixing and the Gribov copies

In both derivations of the KO criterion mentioned above, it was assumed that the gauge fixing is achieved by the usual local gauge fixing condition $\partial_\mu \mathscr{A}^\mu = 0$.
Therefore, the Gribov copies were not taken into account in these derivations.
In order to avoid this problem, global minimization of the gauge fixing functional is necessary. In the lattice gauge theory, the gauge fixing is in principle performed by the global minimization and therefore the results must be free from the Gribov problem (except for the lattice Gribov copies and practical limitations for time and resources)

(ii) compact formulation and including topological configurations

The lattice gauge theory is based on the compact formulation written in terms of the Lie-group valued gauge variables, while the continuum theory is based on the non-compact formulation written in terms of the Lie-algebra valued gauge potentials, although the continuum limit of the lattice gauge theory is expected to essentially reproduce the non-compact continuum theory in the scaling region.
The lattice gauge theory enables to incorporate the topological configurations from the beginning.
According to the lattice study, it has been confirmed that the topological configurations give the dominant contributions for quark confinement.
Whereas the Kugo-Ojima result is based on the canonical operator formalism in the non-compact formulation and do not take into account the topological configurations which are usually incorporated in the path integral formalism.
It is important to take into account the topological objects to bridge the gap between lattice gauge theory and continuum gauge theory.
This is the very reason why we incorporate the topological configuration also in the study of color confinement.



The KO color confinement criterion was reproduced from the viewpoint of the restoration of the RGS, only when the gauge-transformation $\omega(x)$ is linear in $x$, which is referred to as the non-compact case in our paper.
However, the choice for $\omega(x)$ is not unique and there is no sure physical reason to choose the above linear form, since it was just chosen from the analogy in the Abelian gauge theory by the reason it is the simplest form for satisfying the equation (\ref{eq:Laplace}) which is a condition to guarantee the RGS in the Lorenz gauge.
Therefore, the gauge-transformation function $\omega(x)$ can take different forms other than the above specific form, which does not lead to the KO criterion.
Thus we do not consider that the KO criterion is a true criterion for color confinement to be  derivable as the restoration of the RGS even in the Lorenz gauge. Therefore, it is not a surprise to us that KO criterion is not satisfied in the results of numerical simulations on the lattice\cite{Furui-Nakajima,Sternbeck,Aguilar}, while the lattice results support confinement.


Moreover, the gauge-fixed lattice simulations in the last two decades have displayed the high level of sophistication on the issues.
Indeed, they have given the critical reassessment of the entire confinement mechanism that has been caused as a result of the lattice simulations, at the level of the gluon and ghost propagator.
In particular, a series of pivotal results \cite{Boucaud,Bogolubsky,Cucchieri,Sternbeck-Smekal} have dispelled the ``ghost-dominance'' scenario of QCD, which is the intellectual offspring of the KO
formalism.

\section{\label{sec:residual-Lorenz} Extending the residual gauge symmetry restoration in the Lorenz gauge}

In this paper,
we extend the argument of the RGS restoration by including
the topological configurations to see their effects towards color confinement.
This is motivated from the conventional wisdoms mainly obtained after the proposal of the KO criterion based on the lattice gauge theories and the other non-perturbative methods claiming that topological configurations are dominantly responsible for non-perturbative phenomena  such as confinement and chiral symmetry breaking.

\subsection{\label{sec:finite-transformation} A finite version of the generalized gauge transformation and the residual gauge symmetry}

In what follows, we consider the gauge-invariant part of the Lagrangian given by
\begin{align}
 \mathscr{L}_{\text{inv}} &= - \frac{1}{4} \mathscr{F}_{\mu \nu} \cdot \mathscr{F}^{\mu \nu} + \frac{\vartheta}{32 \pi^2} \mathscr{F}_{\mu \nu} \cdot {}^* \mathscr{F}^{\mu \nu} + \mathscr{L}_{\text{matter}} (\varphi , D_\mu \varphi) ,
\end{align}
where $\vartheta$ is a topological angle and ${}^* \mathscr{F}_{\mu \nu}$ is the Hodge dual of the field strength $\mathscr{F}_{\mu \nu}$.
The integer-valued topological charge $Q_P$ is given by the integral  $Q_P=\int d^4x \frac{1}{32 \pi^2} \mathscr{F}_{\mu \nu} \cdot {}^* \mathscr{F}^{\mu \nu}$.

In order to incorporate topological configurations in discussing the restoration of the RGS, we extend the infinitesimal version \cite{Hata-method} of the ``generalized'' local gauge transformation
\footnote{
Usually, the gauge transformation is defined for the gauge fields and matter fields. However, the ``generalized'' gauge transformation is defined also for the unphysical fields $\mathscr{C} , \ \bar{\mathscr{C}} , \ \mathscr{B}$.
}
to a finite one.

Indeed, we introduce \textit{a finite version} of the generalized local gauge transformation by the Lie-group element
\begin{align}
U (x) = e^{i g \omega (x)}, \ \omega (x) = \omega^A (x) T_A
\end{align}
of the gauge group $G$ with the Lie-algebra valued transformation function $\omega (x) = \omega^A (x) T_A$: For the gauge field $\mathscr{A}_\mu (x)$ and the matter field $\varphi (x)$,
 \begin{align}
  \delta_U \mathscr{A}_\mu (x) &:=  U(x) (\mathscr{A}_\mu(x) + i g^{- 1} \partial_\mu) U^\dagger(x) - \mathscr{A}_\mu(x)
\nonumber\\
  &= \Omega_\mu (x) + U (x) \mathscr{A}_\mu (x) U^\dagger (x) - \mathscr{A}_\mu(x),
   \nonumber\\
  \delta_U \varphi (x) &:= U (x) \varphi (x) - \varphi(x),
  \label{eq:gauge-transformation-A,phi}
 \end{align}
 where we have defined a pure gauge form:
  \begin{align}
  \Omega_\mu(x) := i g^{- 1} U(x) \partial_\mu U^\dagger(x) .
 \end{align}
 $\mathscr{L}_{\text{inv}}$ is invariant under this transformation.
 \footnote{
 We can check easily as follows:
\begin{align}
  \delta_U \mathscr{L}_{\text{inv}} &= - \frac{1}{4} \text{tr} [U \mathscr{F}_{\mu \nu} U^\dagger U \mathscr{F}^{\mu \nu} U^\dagger] + \frac{\vartheta}{32 \pi^2} \text{tr} [U \mathscr{F}_{\mu \nu} U^\dagger U {}^* \mathscr{F}^{\mu \nu}U^\dagger ] + \mathscr{L}_{\text{matter}} (U \varphi , U D_\mu U^\dagger U \varphi) - \mathscr{L}_{\text{inv}} \nonumber\\
  &= 0 . \nonumber
\end{align}
 }
  Similarly, for the other fields, we define
 \begin{align}
  \delta_U \mathscr{B} (x) &:= U (x) \mathscr{B}(x) U^\dagger (x) - \mathscr{B}(x), \nonumber\\
  \delta_U \mathscr{C} (x) &:= U (x) \mathscr{C}(x) U^\dagger (x) - \mathscr{C}(x), \nonumber\\
  \delta_U \bar{\mathscr{C}} (x) &:= U (x) \bar{\mathscr{C}}(x) U^\dagger (x) - \bar{\mathscr{C}}(x).
  \label{eq:gauge-transformation-B,C}
 \end{align}
 In order to identify this transformation with the RGS of the Lorenz gauge $\partial^\mu \mathscr{A}_\mu (x) = 0$, the transformation function $U (x)$ should satisfy the following equation almost everywhere for a given $\mathscr{A}_\mu (x)$ satisfying $\partial^\mu \mathscr{A}_\mu (x) = 0$:
 \footnote{
 The residual gauge transformation $U$ as the solution of this equation depends on the given gauge field $\mathscr{A}_\mu$. However, we need the residual gauge transformation $U$ for the perturbative vacuum configuration $\mathscr{A}_\mu \equiv 0$, see \eqref{eq:gfc-perturbative0}.
 }
 \begin{align}
 0 = \partial^\mu \delta_U \mathscr{A}_\mu (x) \ \Leftrightarrow \ 0 = \partial^\mu (\Omega_\mu (x) + U (x) \mathscr{A}_\mu (x) U^\dagger (x)).
 \label{eq:Laplace}
 \end{align}

 We will show that in the next section \ref{sec:topological-configuration} a class of gauge transformation $U$ satisfying this condition indeed exist.
 Therefore we go ahead by assuming the existence of such gauge transformation $U$.
 Notice that we can define another pure gauge form

\begin{align}
  \tilde{\Omega}_\mu(x) := - i g^{- 1} U^\dagger(x) \partial_\mu U(x) .
 \end{align}

It should be remarked that $\Omega_\mu(x)$ and $\tilde{\Omega}_\mu(x)$ agree for the infinitesimal gauge transformation $U (x) = \bm{1} + i g \omega (x)$: $\Omega_\mu(x)=\tilde{\Omega}_\mu(x)=\partial_\mu \omega(x)$.
However, this agreement does not hold in a finite case. The infinitesimal version of the generalized local gauge transformation is given by
\begin{align}
  \delta_\omega \mathscr{A}_\mu (x) &= (\bm{1} + i g \omega (x)) (\mathscr{A}_\mu(x) + i g^{- 1} \partial_\mu) (\bm{1} - i g \omega (x)) - \mathscr{A}_\mu(x) \nonumber\\
  &= i g \omega (x) \mathscr{A}_\mu (x) - i g \mathscr{A}_\mu (x) \omega (x) + \partial_\mu \omega (x) = \mathscr{D}_\mu \omega (x),
 \nonumber\\
  \delta_\omega \varphi (x) &= (\bm{1} + i g \omega (x)) \varphi (x) - \varphi(x) = i g \omega (x) \varphi (x) , \nonumber\\
  \delta_\omega \mathscr{B} (x) &= (\bm{1} + i g \omega (x)) \mathscr{B}(x) (\bm{1} - i g \omega (x)) - \mathscr{B}(x) \nonumber\\
  &= i g \omega (x) \mathscr{B} (x) - i g \mathscr{B} (x) \omega (x) = g (\mathscr{B} \times \omega), \nonumber\\
  \delta_\omega \mathscr{C} (x) &= (\bm{1} + i g \omega (x)) \mathscr{C}(x) (\bm{1} - i g \omega (x)) - \mathscr{C}(x) = g (\mathscr{C} \times \omega), \nonumber\\
  \delta_\omega \bar{\mathscr{C}} (x) &= (\bm{1} + i g \omega (x)) \bar{\mathscr{C}}(x) (\bm{1} - i g \omega (x)) - \bar{\mathscr{C}}(x) = g (\bar{\mathscr{C}} \times \omega).
  \label{eq:infinitesimal_gt}
\end{align}

\subsection{\label{sec:disappearance_NG}Disappearance of the Nambu-Goldstone pole as the restoration of the residual gauge symmetry}
In this subsection, we restrict the gauge transformation to the infinitesimal version in a specific topological sector with a fixed topological charge $Q_P$ since it is enough to consider the infinitesimal gauge transformation in the discussion of the Nambu-Goldstone pole.

In the Lorenz gauge, we can define the ``conserved'' Noether current $\mathscr{J}_{\omega}^\mu$ and the Noether charge $Q_\omega$ associated with the RGS for the generalized gauge transformation \eqref{eq:infinitesimal_gt} according to the standard prescription
\footnote{
The ``conserved'' means that this current $\mathscr{J}_\omega^\mu$ is conserved in the physical subspace $\mathcal{V}_{\text{phys}}$ following the first paper of Hata \cite{Hata-method} i.e. conserved up to the BRST exact term:
\begin{align}
  \partial_\mu \mathscr{J}_\omega^\mu = \delta_{\text{B}} (\cdots) \Rightarrow \ \braket{\text{phys} | \partial_\mu \mathscr{J}_\omega^\mu |\text{phys}} = 0 . \nonumber
\end{align}
Alternatively, we can define the current conserved in the total state space $\mathcal{V}$, which was actually performed in the second paper of Hata, Prog. Theor. Phys. Vol. 69, 1524-1536 (1983). However, the corresponding gauge transformation function $\omega (x ; \mathscr{A}_\mu)$ is explicitly dependent on the gauge field $\mathscr{A}_\mu$ and has non-local character.
}
\begin{align}
  Q_\omega = & \int d^d x \mathscr{J}_{\omega}^0 , \ \mathscr{J}_\omega^\mu = \sum_a \frac{\partial \mathscr{L}}{\partial \partial_\mu \Phi_a} \delta_\omega \Phi_a , \
  \Phi_a = \{\mathscr{A}_\mu , \mathscr{B} , \mathscr{C}, \bar{\mathscr{C}} , \varphi\}.
\end{align}
We discuss the spontaneous breaking of a continuous global symmetry which is associated with the RGS. Remember that the continuous global symmetry to be broken spontaneously is specified by a continuous infinitesimal parameter $\varepsilon$. Therefore, $\delta_\omega$ should be always understood in the form with $\varepsilon$ although we do not write $\varepsilon$ explicitly: $\delta_\omega \rightarrow \ \varepsilon \delta_\omega , \ Q_\omega \rightarrow \ \varepsilon Q_\omega , \ \mathscr{J}_\omega^\mu \rightarrow \ \varepsilon \mathscr{J}_\omega^\mu$, e.g.,
\begin{align}
  \varepsilon \delta_\omega \Phi_a (x) = [i \varepsilon Q_\omega , \Phi_a (x)] , \ \varepsilon Q_\omega = \int d^d x \ \varepsilon \mathscr{J}_\omega^0.
\label{epsilon}
\end{align}
The existence of such a continuous global symmetry $\varepsilon Q_\omega$ and emergence of the global parameter $\varepsilon$ is shown explicitly in the next section after demonstrating the existence of the RGS.

For simplicity, we consider the sector of a single gauge field $\mathscr{A}_\nu^B(y)$ communicating with the conserved current $\mathscr{J}_\omega^\mu(x)$.
This symmetry looks ``spontaneously broken'', because the vacuum expectation value $\braket{0 | [i Q_\omega , \mathscr{A}_\nu (y)] | 0}$ has a non-vanishing value. In what follows, we use a simple notation for the vacuum expectation value: $\braket{\mathcal{O}} := \braket{0 | \mathcal{O} | 0}$.
\begin{align}
  \braket{\delta_\omega \mathscr{A}_\nu^B (y)} = \braket{ [i Q_\omega , \mathscr{A}_\nu^B (y)] } = \braket{\partial_\nu \omega^B (y) + g (\mathscr{A}_\nu \times \omega)^B (y)} = \partial_\nu \omega^B (y) \neq 0 ,
  \label{eq:spontaneously_broken}
\end{align}
where we have used $\braket{\mathscr{A}_\mu} = 0$ which follows from the Lorentz invariance of the vacuum.

The divergence of this current agrees with the generalized gauge transformation of the GF+FP term $\mathscr{L}_{\text{GF+FP}}$ of the Lagrangian due to $\delta_\omega \mathscr{L}_{\text{inv}} = 0$:
\begin{align}
   \partial_\mu \mathscr{J}_\omega^\mu &= \delta_\omega \mathscr{L} = \delta_\omega \mathscr{L}_{\text{GF+FP}}
   = - i \delta_\omega \delta_{\text{B}} \left[\bar{\mathscr{C}}^A \left(\partial^\mu \mathscr{A}_\mu^A + \frac{\alpha}{2} \mathscr{B}^A\right)\right] \nonumber\\
   &= - i \delta_{\text{B}} \delta_\omega \left[\bar{\mathscr{C}}^A \left(\partial^\mu \mathscr{A}_\mu^A + \frac{\alpha}{2} \mathscr{B}^A\right)\right]
   = - i \delta_{\text{B}} \delta_\omega [\bar{\mathscr{C}}^A \partial^\mu \mathscr{A}_\mu^A] = i \delta_{\text{B}} [(\mathscr{D}_\mu \bar{\mathscr{C}})^A] \partial^\mu \omega^A ,
   \label{eq:L_gauge_transform_Lorenz}
 \end{align}
 where we have used the fact that $\delta_{\text{B}}$ and $\delta_\omega$ commute \cite{Kondo-Fukushima}.
 This result holds for any configuration of $\partial_\mu \omega^A$ without restricting to the RGS satisfying \eqref{eq:Laplace} and is independent of $\alpha$.

In the sector of a single gauge field $\mathscr{A}_\nu^B (y)$, we focus on the following Ward-Takahashi (WT) identity
\begin{align}
  i \partial_\mu^x \braket{\mathscr{J}_\omega^\mu (x) \mathscr{A}_\nu^B (y)}
  = \delta^D (x - y) \Braket{\delta_\omega \mathscr{A}_\nu^B (y)} + i \braket{\partial_\mu \mathscr{J}_\omega^\mu (x) \mathscr{A}_\nu^B (y)}.
  \label{eq:WT_original}
\end{align}

We regard the symmetry restoration as the disappearance of the massless Nambu-Goldstone pole ($p^2 = 0$) associated with the spontaneous breaking of the RGS. Therefore, \eqref{eq:spontaneously_broken} is modified into

\begin{align}
\braket{[i Q_\omega , \mathscr{A}_\nu^B (y)]} &= \lim_{p \rightarrow \ 0} \int d^D x \ e^{i p (x - y)} i \partial_\mu^x \braket{\mathscr{J}_\omega^\mu (x) \mathscr{A}_\nu^B (y)} \nonumber\\
&= \braket{\delta_\omega \mathscr{A}_\nu^B (y)} + \lim_{p \rightarrow \ 0} i \int d^D x \ e^{i p (x - y)} \braket{\delta_\omega \mathscr{L} (x) \mathscr{A}_\nu^B (y)}.
\label{eq:A-WI-1_Lorenz}
\end{align}

By using \eqref{eq:L_gauge_transform_Lorenz}, the last term of \eqref{eq:A-WI-1_Lorenz} is cast into
\begin{align}
  &i \int d^D x \ e^{i p (x - y)} \braket{\delta_\omega \mathscr{L} (x) \mathscr{A}_\nu^B (y)} \nonumber\\
  = &- \int d^D x \ e^{i p (x - y)} \partial^\mu \omega^A (x) \braket{\delta_{\text{B}} (\mathscr{D}_\mu \bar{\mathscr{C}})^A (x) \mathscr{A}_\nu^B (y)} \nonumber\\
  = &- \int d^D x \ e^{i p (x - y)} \partial^\mu \omega^A (x) \braket{(\mathscr{D}_\mu \bar{\mathscr{C}})^A (x) \delta_{\text{B}} \mathscr{A}_\nu^B (y)} \nonumber\\
  = &- \int d^D x \ e^{i p (x - y)} \partial^\mu \omega^A (x) \left[\frac{\partial^x_\mu \partial^x_\nu}{\partial_x^2} \delta^{A B} \delta^D (x - y) + \left(g_{\mu \rho} - \frac{\partial^x_\mu \partial^x_\rho}{\partial_x^2}\right) u^{A B} (x - y)\right],
\end{align}
where $u^{A B}$ is the Kugo-Ojima function in the configuration space defined by
\begin{align}
  \braket{0 | (\mathscr{D}_\mu {\mathscr{C}})^A (x) (g \mathscr{A}_\nu \times \bar{\mathscr{C}})^B (y) | 0} = \left(g_{\mu \nu} - \frac{\partial_\mu^x \partial_\nu^x}{\partial_x^2}\right) u^{A B} (x - y).
\end{align}
Then, according to the same argument as that in our previous paper\cite{Kondo-Fukushima}, the general condition for restoration of the RGS in a single gauge field sector $\mathscr{A}_\nu^B (y)$
 is written using
 \begin{align}
   I_\nu^B(y) &:= \lim_{p \rightarrow \  0} \int d^D x \ e^{i p (x - y)} \partial^\mu \omega^A (x) \left(g_{\mu \nu} -  \frac{\partial^x_\mu \partial^x_\nu}{\partial_x^2} \right) (\delta^D (x - y) \delta^{A B} + u^{A B} (x - y)).
\label{eq:lorenz_case_restoration0}
\end{align}
Thus, we arrive at the general condition for restoration of a RGS in the Lorenz gauge given by
\begin{align}
   I_\nu^B(y) &=
   \begin{cases}
     = 0 &\text{restoration} \\
     \neq 0 &\text{no restoration}
   \end{cases}
   .
   \label{eq:lorenz_case_restoration}
 \end{align}
 See Appendix \ref{sec:EOM} for the details on the derivation of this criterion \eqref{eq:lorenz_case_restoration0}.

 In the momentum space, \eqref{eq:lorenz_case_restoration0} reads
 \begin{align}
   I_\nu^B(y) &= \lim_{p \rightarrow \  0} (2 \pi)^D \int d^D k \ e^{- i (p - k) y} \partial^\mu \omega^B (p - k) \left(g_{\mu \nu} - \frac{k_\mu k_\nu}{k^2} \right) (\delta^{A B} + u^{A B} (k)), \nonumber\\
   \partial^\mu \omega^A (q) &:= \int d^D x \ e^{i q x} \partial^\mu \omega^A (x) \ (q := p - k).
   \label{eq:lorenz_case_restoration_p}
 \end{align}

 Once the configuration $\partial_\mu \omega$ is specified from $\{U\}$ which is determined so as to satisfy \eqref{eq:Laplace} for a given initial configuration $\{\mathscr{A}_\mu\}$, the criterion $I_\nu^B (y) = 0$ gives a condition to be satisfied for the KO function $u (k)$ for the restoration. Therefore, we can check whether this criterion is satisfied or not by measuring the KO function $u (k)$ in the whole range of $k$, e.g., using numerical simulations in the frame work of lattice gauge theory.

The KO criterion is reproduced
\begin{align}
  I_\nu^B(y) = \lim_{p \rightarrow \  0}  b_\mu \delta^{A E} \left(g_{\mu \nu} - \frac{p_\mu p_\nu}{p^2}\right) [\delta^{A B} + u^{A B} (p^2)]=0,
  \label{eq:KO2}
\end{align}
only when $\partial^\mu \omega^A (p - k)=\delta^{A E} b_\mu \delta^D (p - k)$ with a specific index $E$ of the Lie algebra.
This case corresponds to the infinitesimal gauge transformation function $\omega(x)$ as the RGS for the Lorenz gauge \eqref{eq:Laplace} which takes a (\textit{non-compact}) form: $\omega(x)$ is linear in $x$:
\begin{align}
 \omega^A (x) = \delta^{A E} b_\mu x^\mu \Rightarrow
 \Omega_\mu^A(x)=\tilde{\Omega}_\mu^A(x) =\partial_\mu \omega^A (x) = \delta^{A E} b_\mu .
\end{align}

In this and only in this case, the condition \eqref{eq:lorenz_case_restoration} is reduced to the Kugo-Ojima color confinement criterion \eqref{eq:KO-criterion}.
In the Lorenz gauge, the condition for restoration of the RGS for this specific choice of $\omega (x)$ agrees with the Kugo-Ojima color confinement criterion (in the the \textit{non-compact} formalism).
This is a remarkable result first shown by Hata \cite{Hata-method}. However, this result can be modified by taking into account the topological defects (in the \textit{compact} case).

\section{\label{sec:topological-configuration} Topological configurations constituting the RGS and contributing to the path integral}
\subsection{$SU (2)$ gauge transformation}

Remember that any element $U$ of the gauge group $SU (2)$ can be expressed as
\begin{align}
  U (x) = \exp \left(i \theta (x) \text{n}^A (x) \frac{\sigma_A}{2}\right) = \cos \frac{\theta (x)}{2} + i \sin \frac{\theta (x)}{2} \text{n}^A (x) \sigma_A \ (A = 1 , 2 , 3) ,
\end{align}
where $\text{n}^A (x)$ is a unit vector $\mathbf{n} (x) \cdot \mathbf{n} (x) = \text{n}^A (x) \text{n}^A (x) = 1$ and $\theta (x)$ is an angle of the rotation around $\mathbf{n} (x)$.
Then, the pure gauge form $\Omega_\mu (x)$ is represented as
\begin{align}
  \Omega_\mu (x) &= i U (x) \partial_\mu U^\dagger (x) \nonumber\\
  &= \frac{\bm{\sigma}}{2} \cdot [\partial_\mu \theta (x) \mathbf{n} (x) + \sin \theta (x) \partial_\mu \mathbf{n} (x) + (1 - \cos \theta (x)) (\partial_\mu \mathbf{n} (x) \times \mathbf{n} (x))] .
\end{align}
It should be noted that $\mathbf{n} , \ \partial_\mu \mathbf{n}$ and $\partial_\mu \mathbf{n} \times \mathbf{n}$ constitute the orthogonal bases of $su (2)$. For example, a specific choice of $\mathbf{n}$ leads to
\begin{align}
    \text{n}^A = \frac{x_A}{r}, \ \partial_0 \text{n}^A = 0 , \ (\partial_0 n \times n)^A = 0 , \  \partial_j \text{n}^A = \frac{\delta_{j A} r^2 - x_j x_A}{r^3} , \ (\partial_j n \times n)^A = \frac{\varepsilon_{j A k} x_k}{r^2} ,
\end{align}
where we have defined the spatial radius $r = \sqrt{x_1^2 + x_2^2 + x_3^2}$.

\subsection{Witten Ansatz}

We have considered the restoration condition by assuming the existence of the RGS in the section \ref{sec:residual-Lorenz}.
In order to examine that the gauge transformation matrix $U$ satisfying \eqref{eq:Laplace} indeed exists, we focus on the $SU(2)$ Yang-Mills theory in $D = 4$ Euclidean space and adopt, as one example, the following Ansatz for the $SU(2)$ gauge field $\mathscr{A}_\mu^A (A=1,2,3)$ with the cylindrical symmetry which was invented by Witten \cite{Witten1977}:
\begin{align}
  -\mathscr{A}_4^A (x) &= \frac{x_A}{r} A_0 (r , t) , \nonumber\\
  -\mathscr{A}_j^A (x) &= \frac{x_j x_A}{r^2} A_1 (r , t) + \frac{\delta_{j A} r^2 - x_j x_A}{r^3} \varphi_1 (r , t)  + \frac{\varepsilon_{j A k} x_k}{r^2} [1 + \varphi_2 (r , t)] \ (j = 1 , 2 , 3),
  \label{eq:Witten_Ansatz}
\end{align}
where $A_0, A_1, \varphi_1$ and $\varphi_2$ are unknown Ansatz functions with the cylindrical symmetry depending only on the Euclidean time $t$ and the spatial radius $r$.

The Witten Ansatz was originally introduced to find the multi-instanton solution of the self-dual equation $\mathscr{F}_{\mu \nu} = \pm {}^* \mathscr{F}_{\mu \nu}$ in the $D = 4$ Euclidean space $(t , x_1 , x_2 , x_3)$. In this paper, we use this Ansatz to simplify the equation \eqref{eq:Laplace}, namely, the Gribov equation \eqref{gfc} for finding the RGS, rather than finding the multi-instanton solution.

An advantage of using the Witten Ansatz is to map the $SU (2)$ Yang-Mills theory to another theory defined in the two-dimensional Euclidean space $(t , r)$ which is at most an Abelian gauge theory since the non-Abelian group $SU (2)$ structure was fixed by choosing the field $\mathbf{n} (x) \ \left(\text{n}^A (x) = \frac{x^A}{r}\right)$ constituting the three orthogonal bases $\mathbf{n} (x) , \ \partial_\mu \mathbf{n}$ and $\partial_\mu \mathbf{n} (x) \times \mathbf{n} (x)$ of $SU (2)$:
\begin{align}
  \mathscr{A}_\mu^A (x) = \text{n}^A (x)
  \begin{cases}
    A_0 (r , t) & (\mu = 0)\\
    \frac{x_j}{r} A_1 (r, t) & (\mu = j)
  \end{cases}
  + \partial_\mu \text{n}^A (x) \varphi_1 (r , t) + (\partial_\mu \mathbf{n} \times \mathbf{n})^A \varphi_2 (r , t) .
\end{align}

\color{black}

The ordinary instantons with the \textit{four-dimensional spacetime spherical symmetry}  in $D = 4$  Euclidean spacetime are generally considered to be irrelevant for confinement.
However, it is important to recall that instantons as solutions of the self-dual equation respect various symmetries.
The instantons we consider in this paper to be relevant to RGS and confinement are those with a certain \textit{three-dimensional spatial spherical symmetry}, which corresponds to the cylindrical symmetry (or axial symmetry around the Euclidean time axis) from the perspective of four-dimensional spacetime.


The incorporation of spacetime or space symmetries to the Yang-Mills field and the associated dimensional reduction of spacetime on which theory is defined are discussed in detail in \cite{Forgacs-Manton} and section 4.3 of \cite{Manton}.
In particular, it is proven that if the $D = 4$ $SU (2)$ Yang-Mills field is restricted to the configuration with the three-dimensional spatial symmetry $SO(3)$, the $D = 4$ $SU (2)$ Yang-Mills theory reduces to the $D=2$ $U(1)$ gauge-scalar theory which is indeed obtained by applying the Witten Ansatz to the original Yang-Mills field.
This dimensional reduction occurs because the three-dimensional spatial rotation $SO(3)$ of the Yang-Mills field is undone by the $SU(2)$ gauge transformation.

We consider that the restoration of the RGS is related to confinement and that topological configurations contribute to confinement.
We suppose that such topological configurations in the pure Yang-Mills theory are given by a class of instantons with a certain space symmetry, from which the important topological configurations such as magnetic monopoles or vortices are derived.
\color{black}
It is well known that vortices and magnetic monopoles play the dominant role in confinement, and our argument in this paper is consistent with this fact.
\footnote{
\color{black}
In the $D=2$ $U(1)$ gauge-scalar theory obtained by such dimensional reduction, it is known that vortex solutions exist, see section 7.14.3 of \cite{Manton}.
The restoration of the RGS through this dimensional reduction and its relation to lower-dimensional topological soliton solutions giving a finite four-dimensional Euclidean action will be discussed in more detail in a subsequent paper in preparation.
In fact, it is shown that the $U(1)$ RGS is restored due to the condensation of such vortices.
}

\color{black}

Restricting the Yang-Mills field to the above class corresponds to focusing the instanton configuration carrying the minimal topological charge.  As a result, this configuration not only makes the four-dimensional action (integral) finite, but also gives a nontrivial minimal action and therefore can be considered one of the most effective and important configurations for the path integral.

The relation between the Witten Ansatz and the 't Hooft Ansatz \cite{tHooft-tensor} is obtained by taking
\begin{align}
  A_0 = F (r , t) r , \ A_1 = - F (r , t) t , \ \varphi_1 = - F (r , t) r t , \ 1 + \varphi_2 = F (r , t) r^2,
\end{align}
as follows
\begin{align}
  \mathscr{A}_4^A (x) &= - x_A F (r , t) ,\nonumber\\
  \mathscr{A}_j^A (x) &= \frac{x_j x_A}{r^2} F (r , t) t + \frac{\delta_{j A} r^2 - x_j x_A}{r^2} F (r , t) t - \varepsilon_{j A k} x_k F (r , t) \nonumber\\
  &= F (r , t) \delta_{j A} t + F (r , t) \varepsilon_{A j k} x_k \nonumber\\
  \Rightarrow \ \mathscr{A}_\mu^A (x) &= F (r , t) \eta^A_{\mu \nu} x_\nu
\end{align}
with the symbol $\eta^A_{\mu \nu}$ defined by
\begin{align}
    \eta^A_{\mu \nu} &= \varepsilon_{A \mu \nu 4} + \delta_{A \mu} \delta_{\nu 4} - \delta_{\mu 4} \delta_{A \nu} =
 \begin{cases}
     \varepsilon_{A \mu \nu} &(\mu , \nu = 1 , 2 , 3) \\
     \delta_{A \mu} &(\nu = 4) \\
    -\delta_{A \nu} &(\mu = 4)
  \end{cases}
  \nonumber\\
  &= - \eta^A_{\nu \mu} .
\end{align}

In what follows, we use the notation: $\partial_0 := \frac{\partial}{\partial t}$ and $\partial_1 := \frac{\partial}{\partial r}$.
The field strength $\mathscr{F}_{\mu\nu}^A$ is expressed using the Ansatz (\ref{eq:Witten_Ansatz}) as follows.
\begin{align}
  - \mathscr{F}_{4 j}^A = &  \frac{x_j x_A}{r^2} F_{0 1}  +  \frac{\delta_{j A} r^2 - x_j x_A}{r^3} D_0 \varphi_1 + \frac{\varepsilon_{j A k} x_k}{r^2} D_0 \varphi_2 ,  \nonumber\\
   - \frac{1}{2} \varepsilon_{jk \ell} \mathscr{F}_{k \ell}^A  = &  \frac{x_j x_A }{r^2} \frac{\varphi_1^2 + \varphi_2^2-1}{r^2}  -  \frac{\delta_{j A} r^2 - x_j x_A }{r^3} D_1 \varphi_2 + \frac{\varepsilon_{j A k} x_k}{r^2} D_1 \varphi_1 ,
   \label{F}
\end{align}
where we have defined
\begin{align}
  F_{\mu \nu} &:= \partial_\mu A_\nu - \partial_\nu A_\mu , \ D_\mu \varphi_a := \partial_\mu \varphi_a + \varepsilon_{a b} A_\mu \varphi_b \ (\mu , \nu = 0 , 1 , \ a , b = 1 , 2) .
\end{align}
Therefore, the Yang-Mills Lagrangian density is rewritten in terms of the Ansatz functions into
\begin{align}
  \mathscr{L}_{\text{YM}} &= \frac{1}{4} \mathscr{F}_{\mu \nu}^A (x) \mathscr{F}_{\mu \nu}^A (x) + \frac{\vartheta}{32 \pi^2} \mathscr{F}_{\mu \nu} {}^* \mathscr{F}_{\mu \nu} \nonumber\\
  &= \frac{1}{2} (\mathscr{F}_{4 j}^A)^2 + \frac{1}{2} \left(\frac{1}{2} \varepsilon_{jk \ell} \mathscr{F}_{k \ell}^A\right)^2 + \frac{\vartheta}{32 \pi^2} \mathscr{F}_{\mu \nu} {}^* \mathscr{F}_{\mu \nu} \nonumber\\
  &= \frac{1}{r^2} D_\mu \varphi_a D_\mu \varphi_a + \frac{1}{4} F_{\mu \nu} F_{\mu \nu} + \frac{1}{2 r^4}(1 - \varphi_a \varphi_a)^2  + \frac{\vartheta}{16 \pi^2 r^2} \varepsilon_{\mu \nu} F_{\mu \nu},
  \label{YM-L}
\end{align}
where we have ignored the total derivative of the topological term \cite{Witten1977}. This implies that the $SU (2)$ Yang-Mills theory with a topological term in $D=4$ dimensional Euclidean space under the Witten Ansatz reduces to the Abelian $U(1)$ gauge-scalar theory with the corresponding topological term in $D=2$ dimensional Euclidean space with the coordinates $(r,t)$  (with a curved metric $g_{\mu\nu}= r^{- 2} \delta_{\mu\nu}$).

In order to understand the meaning of the Witten Ansatz (\ref{eq:Witten_Ansatz}) and the correspondence between the $D=4$ $SU(2)$ Yang-Mills theory and the $D=2$ $U(1)$ gauge-scalar model furthermore,
we perform a specific $SU(2)$ gauge transformation with the cylindrical symmetry expressed by
\begin{align}
U = \exp \left(i \theta (r , t) \frac{x_A}{r} \frac{\sigma_A}{2}\right) = \cos \frac{\theta (r , t)}{2} + i \sin \frac{\theta (r , t)}{2} \frac{x_A}{r} \sigma_A \in SU(2) .
\label{gt-U1}
\end{align}
Under this transformation (\ref{gt-U1}), the gauge field $\mathscr{A}_\mu$ is transformed as
\begin{align}
  \mathscr{A}_4 \rightarrow \ &\mathscr{A}_4^\prime =U \mathscr{A}_4 U^\dagger + i U \partial_0 U^\dagger 
 =- \frac{\sigma_A}{2} \frac{x_A}{r} \left(A_0 - \partial_0 \theta\right) , \nonumber\\
   \mathscr{A}_j \rightarrow \ &\mathscr{A}_j^\prime  =U \mathscr{A}_j U^\dagger + i U \partial_j U^\dagger \nonumber\\
  = &- \frac{\sigma_A}{2} \Biggl[ \frac{x_j x_A}{r^2} (A_1 - \partial_1 \theta)  +   \frac{\delta_{j A} r^2-x_j x_A}{r^3} (\varphi_1 \cos \theta + \varphi_2 \sin \theta)  \nonumber\\
  &\quad \quad\quad+ \frac{\varepsilon_{j A k} x_k}{r^2} (1  - \varphi_1 \sin \theta + \varphi_2 \cos \theta ) \Biggr],
  \label{U1-gt}
\end{align}
where we have used
\begin{align}
  &\frac{x_A}{r} \sigma_A \frac{x_B}{r} \sigma_B = \frac{x_A}{r} \frac{x_B}{r} (\delta_{A B} I + i \varepsilon_{A B C} \sigma_C) = I , \nonumber\\
  &\partial_j =\frac{\partial}{\partial x^j} =\frac{\partial r}{\partial x^j} \frac{\partial}{\partial r} =\frac{x_j}{r} \frac{\partial}{\partial r}  = \frac{x_j}{r} \partial_1 .
\end{align}
Then, the gauge transformation by the element (\ref{gt-U1}) is equivalent to a \textit{finite} $U(1)$ gauge transformation in $D = 2$ $U (1)$ gauge-scalar model as explicitly noted by Actor \cite{Actor79}:
\begin{align}
  A_\mu \rightarrow \ A_\mu^\prime = A_\mu - \partial_\mu \theta , \
  \begin{pmatrix}
    \varphi_1 \\
    \varphi_2
  \end{pmatrix}
  \rightarrow \
  \begin{pmatrix}
    \varphi_1^\prime \\
    \varphi_2^\prime
  \end{pmatrix}
=  \begin{pmatrix}
    \cos \theta & \sin \theta \\
    - \sin \theta & \cos \theta
  \end{pmatrix}
  \begin{pmatrix}
    \varphi_1 \\
    \varphi_2
  \end{pmatrix}
  .
  \label{eq:gauge_transformation_Witten_Ansatz}
\end{align}
This result indicates that the group element (\ref{gt-U1}) corresponds to a compact Abelian subgroup (maximal torus) $U(1)$  of the original non-Abelian compact group $SU(2)$.

If we restrict to the case of the ``cylindrical symmetry'' by adopting
\begin{align}
  \text{n}^A = \frac{x_A}{r}, \ \theta = \theta (r , t) \ (r := \sqrt{x_1^2 + x_2^2 + x_3^2}) ,
\end{align}

we reproduce the Witten Ansatz for the pure gauge form in the consistent way:
\begin{align}
  \Omega_0 (x) := i U \partial_0 U^\dagger &= \frac{\sigma_A}{2} \frac{x_A}{r} \frac{\partial \theta}{\partial t} , \nonumber\\
  \Omega_j (x) := i U \partial_j U^\dagger &= \frac{\sigma_A}{2} \left\{  \frac{x_j x_A}{r^2} \frac{\partial \theta}{\partial r}  + \frac{\delta_{j A} r^2 - x_j x_A}{r^3} \sin \theta - \frac{\varepsilon_{j A k} x_k}{r^2} [1 - \cos \theta]  \right\} .
  \label{eq:Omega_Witten_Ansatz}
\end{align}
Another pure gauge form $\tilde{\Omega}_\mu (x)$ is given by replacing $\theta \rightarrow \ - \theta$ and changing the overall sign:
\begin{align}
  \tilde{\Omega}_0 (x) := - i U^\dagger \partial_0 U &= \frac{\sigma_A}{2} \frac{x_A}{r} \frac{\partial \theta}{\partial t} , \nonumber\\
  \tilde{\Omega}_j (x) := - i U^\dagger \partial_j U &= \frac{\sigma_A}{2} \left\{\frac{x_j x_A}{r^2} \frac{\partial \theta}{\partial r}  + \frac{\delta_{j A} r^2 - x_j x_A}{r^3} \sin \theta + \frac{\varepsilon_{j A k} x_k}{r^2} [1 - \cos \theta]  \right\} .
  \label{eq:Omega_tilde_Witten_Ansatz}
\end{align}
This pure gauge form corresponds to the Ansatz functions in the Witten Ansatz:
\begin{align}
  A_0 = \partial_0 \theta (r , t) , \ A_1 = \partial_1 \theta (r , t) , \ \varphi_1 = \sin \theta (r , t) , \ \varphi_2 = - \cos \theta (r , t).
\end{align}






\subsection{Constituting the residual gauge symmetry}

By using
\begin{align}
  - \partial_0 \mathscr{A}_4^A = &\frac{x_A}{r} \partial_0 A_0 , \nonumber\\
  - \partial_{j} \mathscr{A}_j^A = &\frac{2 \delta_{j A} x_j - 3 x_A - x_j \delta_{j A}}{r^3} \varphi_1 +  \frac{x_A}{r}  \partial_1 A_1 +  \frac{3 x_A r^2 + x_A r^2 - 2 r^2 x_A}{r^4} A_1 \nonumber\\
  = &\frac{x_A}{r^3} [r^2 \partial_1 A_1 + 2 r A_1 - 2 \varphi_1]
 = \frac{x_A}{r^3} [\partial_1 (r^2 A_1) - 2 \varphi_1]  ,
\end{align}
the Lorenz gauge fixing condition $0=\partial_\mu \mathscr{A}_\mu^A = \partial_0 \mathscr{A}_4^A+\partial_{j} \mathscr{A}_j^A$ reduces to
\begin{align}
  \partial_\mu (r^2 A_\mu(r , t)) = 2 \varphi_1(r , t).
  \label{original-gfc}
\end{align}

In order for the $U (1)$ gauge transformation \eqref{eq:gauge_transformation_Witten_Ansatz} obtained from the original $SU (2)$ gauge transformation (\ref{gt-U1}) to become the RGS of the Lorenz gauge fixing $0=\partial_\mu \mathscr{A}^\prime{}_\mu^A$, $\theta$ must satisfy
\begin{align}
  &\partial_\mu [r^2 (A_\mu(r , t) - \partial_\mu \theta(r , t))] = 2 (\cos \theta(r , t) \varphi_1(r , t) + \sin \theta(r , t) \varphi_2(r , t)) ,
\end{align}
which reduces using (\ref{original-gfc}) to
\begin{align}
\partial_\mu (r^2 \partial_\mu \theta (r , t))
  =  2(1-\cos \theta (r , t)) \varphi_1 (r , t) -2 \sin \theta (r , t) \varphi_2 (r , t) \ (\mu = 0 , 1)  .
\label{gfc}
\end{align}
This is nothing but the Gribov equation, see Appendix of \cite{Gribov}.
Thus, the restricted gauge transformation $U = \exp \left(i \theta (r , t) \frac{x_A}{r} \frac{\sigma_A}{2}\right)$ given by (\ref{gt-U1}) becomes the RGS of the Lorenz gauge,  only when $\theta$ is a solution of the Gribov equation (\ref{gfc}) for a given original  gauge field configuration $\mathscr{A}_j^A$ specified by $\varphi_1$ and $\varphi_2$.

For our purpose of discussing spontaneous symmetry breaking, it is enough to focus on the vacuum configuration $\mathscr{A}_\mu \equiv 0$.
For this perturbative vacuum $\mathscr{A}_\mu = 0$ corresponding to $A_0 = A_1 = \varphi_1 = 0$ and $\varphi_2 = - 1$, \eqref{gfc} reduces to
\begin{align}
  \partial_\mu (r^2 \partial_\mu \theta (r , t)) - 2 \sin \theta (r , t) = 0.
  \label{eq:gfc-perturbative0}
\end{align}
This is further simplified in the $t$-independent case:
\begin{align}
  \partial_1 (r^2 \partial_1 \theta (r)) - 2 \sin \theta (r) = 0.
  \label{eq:gfc-perturbative}
\end{align}
In order to linearize the equation \eqref{eq:gfc-perturbative} around $\theta = \theta_0$, we substitute $\theta (r) = \theta_0 + \delta (r)$. Then, \eqref{eq:gfc-perturbative} reduces to the second order linear differential equation of the Euler type:
\begin{align}
  r^2 \delta^{\prime \prime}(r) + 2 r \delta^{\prime}(r) - 2 \cos \theta_0 \delta(r) = 2 \sin \theta_0,
\end{align}
where the prime denotes the differentiation with respect to $r$. Therefore, the asymptotic solution for $r \rightarrow \ 0$ or $r \rightarrow \ \infty$ is given by
\begin{align}
  \theta (r) \sim
  \begin{cases}
    \theta_0 - \tan \theta_0 + C_1 r^{\frac{1}{2}(- 1 + \sqrt{1 + 8 \cos \theta_0})} \hspace{10ex} (1 + 8 \cos \theta_0 > 0 , \ r \rightarrow \ 0) \\
    \theta_0 - \tan \theta_0 + C_2 r^{- \frac{1}{2}} \cos \left(\frac{1}{2}\sqrt{|1 + 8 \cos \theta_0|} \ln r + \alpha\right) \\
    \hspace{44ex} (1 + 8 \cos \theta_0 < 0 , \ r \rightarrow \ \infty)
  \end{cases}
  ,
\end{align}
which specifies the global $U (1)$ symmetry.
The existence of the solution $\theta (r)$ of the non-linear Gribov equation \eqref{eq:gfc-perturbative} for any $r$ can be understood by rewriting it into the equation of motion for the pendulum in the uniform gravitational field by introducing the time variable through $\tau = \ln r$:
\begin{align}
  \ddot{\theta} (\tau) + \dot{\theta} (\tau) - \sin \theta (\tau) = 0.
\end{align}
Then we find $\theta_0 = 0$ for $r \rightarrow \ 0 \ (\tau \rightarrow \ - \infty)$ and $\theta_0 = \pm \pi$ for $r \rightarrow \ \infty \ (\tau \rightarrow \ + \infty)$.
Therefore the solution for $\theta (r)$ indeed exists as shown by Gribov \cite{Gribov} where the solution has the non-trivial topology $Q_P \not=0$. We find that this solution has a global parameter: $C_1$ or $C_2$ which plays the role of the global parameter $\varepsilon$ in (\ref{epsilon}).
The existence of the solution in more general case \eqref{gfc} was also examined, see \cite{Actor79} and references therein. Thus, the gauge transformation $U$ satisfying \eqref{eq:Laplace} indeed exists and we can discuss the Nambu-Goldstone pole associated with the global $U (1)$ symmetry.

\subsection{procedure}
The procedure of checking our criterion for the restoration of the RGS is as follows.
\begin{enumerate}
    \item Evaluate the KO function $u (x)$ in the original Yang-Mills theory written in terms of the fields $\mathscr{A}_\mu, \mathscr{B}, \mathscr{C}, \bar{\mathscr{C}}$.
    \item First, confirm that the equation for the RGS in the Lorenz gauge \eqref{eq:Laplace}: $\partial^\mu \Omega_\mu (x) = 0$ is satisfied to find the solution $U (x)$.
    [By adopting the Witten Ansatz, $\mathscr{A}_\mu$ is transferred to $A_0 , A_1 , \varphi_1 , \varphi_2$ and the equation \eqref{eq:Laplace} is reduced to the Gribov equation \eqref{gfc}. Then solve the Gribov equation \eqref{gfc} to find the solution $\theta$.]
    \item Construct $\partial_\mu \omega$ from the pure gauge $\Omega = i g^{- 1} U (x) \partial_\mu U^\dagger (x)$ using the obtained $\theta (x)$ and $\mathbf{n} (x)$ in the previous step through $U (x) = \exp \left(i \theta (x) \mathbf{n} (x) \cdot \frac{\bm{\sigma}}{2}\right)$.
    \item Substitute $u (x)$ and $\partial_\mu \omega (x)$ into the criterion \eqref{eq:lorenz_case_restoration0} (or \eqref{eq:lorenz_case_restoration_p} after the Fourier transform) and check whether $I_\nu^B = 0$ or not.
\end{enumerate}

\subsection{Giving a finite Euclidean action and contributing to the path integral}
Since the integration measure is transformed as
\begin{align}
  d^4 x &= d t d^3 x = 4 \pi d t d r \ r^{2} ,
\end{align}
we can express the $D = 4$ Euclidean Yang-Mills action $S_{\text{YM}} = \int d^4 x \ \mathscr{L}_{\text{YM}}$ as

\begin{align}
  S_{\text{YM}} = 4 \pi \int_{- \infty}^\infty d t \int_0^\infty d r \ \Biggl[\frac{1}{2} D_\mu \varphi_a D_\mu \varphi_a &+ \frac{1}{8} r^2 F_{\mu \nu} F_{\mu \nu} \nonumber\\
  &+ (1 - \varphi_a \varphi_a)^2 \frac{1}{4 r^2} + \frac{\vartheta}{16 \pi^2} \varepsilon_{\mu \nu} F_{\mu \nu}\Biggr] .
  \label{eq:YM2_action}
\end{align}
We must choose the Ansatz functions $A_0$, $A_1$, $\varphi_1$ and $\varphi_2$ such that
the topological configuration gives a finite Euclidean action $S_{\text{YM}}<\infty$ and contributes to the path integral.

For the time-independent topological configuration (e.g., as realized by restricting to the $D = 3$ case), in order for this Euclidean action integral to give a finite value (without integration over $t$ variable), we find that the Ansatz functions must satisfy the following boundary conditions at $r = 0$ and $r = \infty$:
\begin{align}
  &
  \begin{cases}
    D_\mu \varphi_a = \mathcal{O} \left(\left(\frac{1}{r}\right)^{\frac{1}{2} + \delta_1}\right) & (r \rightarrow \ \infty , \ 0 < \delta_1)\\
    \varphi_a \varphi_a = 1 + \mathcal{O} (r^{\frac{1}{2} + \delta_2})
    , \ D_\mu \varphi_a = \mathcal{O} (r^{- \frac{1}{2} + \delta_3})
    & (r \rightarrow \ 0 , \ 0 < \delta_2 , \delta_3)
  \end{cases}
  , \nonumber\\
  &F_{\mu \nu} =
  \begin{cases}
    \mathcal{O} \left(\left(\frac{1}{r}\right)^{\frac{3}{2} + \delta_4}\right) & (r \rightarrow \ \infty , \ 0 < \delta_4)\\
    \mathcal{O} (r^{- \frac{3}{2} + \delta_5}) & (r \rightarrow \ 0 , \ 0 < \delta_5)
  \end{cases}
  .
  \label{eq:finite_condition}
\end{align}
For $D = 3$, indeed, we find that this condition \eqref{eq:finite_condition} is not satisfied due to $\varphi_a \varphi_a = 0$ for one magnetic monopole of the 'tHooft-Polyakov type\cite{tP-monopole} in the zero size limit or one magnetic monopole of the Wu-Yang type\cite{WY-monopole} specified by the Ansatz functions:
\begin{align}
  A_0 = A_1 = \varphi_1 = \varphi_2 = 0 \ \Leftrightarrow \ \mathscr{A}_4^A (x) = 0 , \ \mathscr{A}_j^A (x) = \frac{\varepsilon_{j A k} x_k}{r^2} .
\end{align}

For time-dependent  topological configurations ($D = 4$ case), on the other hand,
we find that one meron of the Alfaro-Fubini-Furlan type\cite{AFF-meron} specified by the Ansatz functions:
\begin{align}
  A_0 = \frac{r}{x^2} , \ A_1 = - \frac{t}{x^2} , \ \varphi_1 = - \frac{r t}{x^2} , \ \varphi_2 = - \frac{t^2}{x^2} ,
\end{align}
gives a divergent Euclidean action as is well known, which is also checked by performing the full integration in \eqref{eq:YM2_action}.
These topological configuration are taken in our previous paper\cite{Kondo-Fukushima}, although they give the divergent Euclidean action.

For $D = 3$, it is known \cite{Nishino2018} that the magnetic monopole of 'tHooft-Polyakov type \cite{tP-monopole} with non-vanishing size gives a finite Euclidean action specified by the Ansatz function:
\begin{align}
A_0 = A_1 = \varphi_1 = 0 , \ \varphi_2 \neq 0.
\end{align}
For $D = 4$, we know an example of the topological configuration with a finite Euclidean action: the BPST instanton \cite{BPST-instanton} with a non-vanishing size $\lambda$ \cite{Actor79} specified by the Ansatz functions:
\begin{align}
  &A_0 = F (r , t) r , \ A_1 = - F (r , t) t , \ \varphi_1 = - F (r , t) r t , \ 1 + \varphi_2 = F (r , t) r^2 , \nonumber\\
  &F(r , t) := \frac{2}{r^2 + t^2 + \lambda^2} .
  \label{eq:Instanton_Witten}
\end{align}
Thus, we can choose the topological configuration which constitutes the RGS and gives a finite Euclidean action to contribute to the path integral.

\section{\label{sec:conclusion}Conclusion and discussion}

In this paper, we have reexamined the restoration of the RGS in the Lorenz gauge in detail by properly including the topological configurations towards understanding color confinement.
Indeed, we have given an extended criterion (21) for the restoration of the RGS applicable to any finite gauge transformation to incorporate topological effects, which reduces to the Kugo-Ojima criterion in the special case without topological effects adopted in Hata\cite{Hata-method}.

For this purpose, we have shown the existence of the RGS in the Lorenz gauge obtained from a finite gauge transformation realizing topological configurations with the help of the Witten Ansatz for the SU(2) Yang-Mills theory.
Moreover, the relevant topological configurations give a finite Euclidean action to give non-trivial contribution to the path integral, in sharp contrast to some topological configurations considered in the previous paper\cite{Kondo-Fukushima} giving an infinite Euclidean action due to the short distance singularity of the topological defect.

The obtained criterion (23) indicates that the information of the KO function $u(k)$ in the whole momentum range $k$ is needed to judge the restoration when topological configurations are taken account into considerations of the RGS.
It is of great interest to perform the verification of the new criterion using the numerical simulations on the lattice under the effects of topological configurations.
In other words, it would be possible to check which topological configurations are most responsible to the restoration by checking whether the new criterion is satisfied or not, once the KO function $u(k)$ is measured and known in the whole momentum range.

We have not yet shown at present that our criterion for the restoration has an exact relationship to color confinement.
We are just based on the general belief that confinement will be realized in the disordered phase where all the internal symmetries are unbroken, implying that the RGS is also restored.
Therefore, even in the Lorenz-Landau gauge the true criterion for color confinement and the relationship between the restoration and the confinement is still an open question.
In light of analysis, however, the Kugo-Ojima criterion $u(k=0)=-1$ would not be a true criterion for color confinement.
This is already a consensus in the community as reviewed in section 2.
This is a reason why we reconsider this issue in various gauge fixing conditions by taking into account the topological effects.
We consider this investigation as a first step to find the true confinement criterion.


Color confinement must be a gauge-independent phenomenon. Hence we wish to extend the restoration condition of RGS to any other gauges. The topological configuration considered in this paper can be applied to any other gauge.
To give the complete discussion for confinement, it is necessary to consider interactions among various fields, including different kinds of fields such as the ghost and quark field. Therefore, we hope to investigate the restoration conditions of RGS in sectors of all the fields in a subsequent paper.

\section*{Acknowledgment}
This work was supported by Grant-in-Aid for Scientific Research, JSPS KAKENHI Grant Number (C) No.23K03406.
N.F. was supported by JST, the establishment
of university fellowships towards the creation of science
technology innovation, Grant Number JPMJFS2107 and Grant Number JPMJSP2109.
The authors would like to express sincere thanks to Professor Taichiro Kugo for valuable discussions which greatly help them to revise this paper.

\appendix

\section{\label{sec:EOM}The criterion for the restoration of residual gauge symmetry in a sector of the single gauge field in the Lorenz gauge}


In the previous paper \cite{Kondo-Fukushima}, we derived the criterions corresponding to \eqref{eq:lorenz_case_restoration0}
using the Schwinger-Dyson equation of the anti-ghost field $0 = \int d \mu \frac{\delta}{\delta \bar{\mathscr{C}}^A (y)} (e^{i S} F)$.
However, this method is only applicable to the single gauge field sector, as we used the relation $0 = \frac{\delta e^{i S}}{\delta (- \bar{\mathscr{C}}^A (y))} = e^{i S} \partial^\mu \delta_\text{B} \mathscr{A}_\mu^A (y)$. In this Appendix, we give a more general derivation of the criterion which is applicable to arbitrary sectors $\Phi (y)$ by using the Schwinger-Dyson equations of the ghost field $0 = \int d \mu \frac{\delta}{\delta c^\ell (x)} (e^{i S} F)$ and the NL field $0 = \int d \mu \frac{\delta}{\delta B^a (x)} (e^{i S} F)$.

We consider the WT identity
\begin{align}
 i \partial_\mu^x \braket{\mathscr{J}_\omega^\mu (x) \Phi (y)} = \delta^D (x - y) \Braket{\delta_\omega \Phi (x)} + i \braket{\partial_\mu \mathscr{J}_\omega^\mu (x) \Phi (y)}.
 \label{eq:A-WI}
\end{align}
Using $\partial_\mu \mathscr{J}_\omega^\mu = \delta_\omega \mathscr{L}$, the Fourier transformation of \eqref{eq:A-WI} reads
\begin{align}
 \int d^D x \ e^{i p (x - y)} i \partial_\mu^x \braket{\mathscr{J}_\omega^\mu (x) \Phi (y)} = \braket{\delta_\omega \Phi (y)} + i \int d^D x \ e^{i p (x - y)} \braket{\delta_\omega \mathscr{L} (x) \Phi (y)}.
 \label{eq:A-WI2}
\end{align}

In what follows, we rewrite the last term of \eqref{eq:A-WI2}. Since $\delta_\omega \mathscr{L}$ can be generally written in the BRST exact form, $\delta_\omega \mathscr{L} = i \delta_{\text{B}} F$, $F$ becomes Grassmann odd. Then, the last term of \eqref{eq:A-WI2} is
\begin{align}
 i \int d^D x \ e^{i p (x - y)} \braket{\delta_\omega \mathscr{L} (x) \Phi (y)} &= - \int d^D x \ e^{i p (x - y)} \braket{\delta_{\text{B}} F (x) \Phi (y)} \nonumber\\
 &= - \int d^D x \ e^{i p (x - y)} \braket{F (x) \delta_{\text{B}} \Phi (y)},
 \label{eq:A-WI3}
\end{align}
where we have used $\braket{\delta_{\text{B}} (\cdots)} = 0$.

In the Lorenz gauge, $F (x) = (\mathscr{D}_\mu \bar{\mathscr{C}} (x))^A \partial^\mu \omega^A (x)$, therefore, \eqref{eq:A-WI3} is reduced to
\begin{align}
 - \int d^D x \ e^{i p (x - y)} \braket{\delta_\omega \mathscr{L} (x) \Phi (y)} = \int d^D x \ e^{i p (x - y)} \partial^\mu \omega^A (x) \braket{(\mathscr{D}_\mu \bar{\mathscr{C}} (x))^A \delta_{\text{B}} \Phi (y)}.
\end{align}
Now we focus on the expectation value in the integrand. Decomposing the longitudinal part and transverse part,
\begin{align}
 &\braket{(\mathscr{D}_\mu \bar{\mathscr{C}} (x))^A \delta_{\text{B}} \Phi (y)} \nonumber\\
 = &\frac{\partial^x_\mu \partial^x_\rho}{\partial_x^2} \braket{(\mathscr{D}^\rho \bar{\mathscr{C}})^A (x) \delta_{\text{B}} \Phi (y)} + \left(g_{\mu \rho} - \frac{\partial^x_\mu \partial^x_\rho}{\partial_x^2}\right) \braket{(\mathscr{D}^\rho \bar{\mathscr{C}})^A (x) \delta_{\text{B}} \Phi (y)} \nonumber\\
 = &\frac{\partial^x_\mu \partial^x_\rho}{\partial_x^2} \braket{(\mathscr{D}^\rho \bar{\mathscr{C}})^A (x) \delta_{\text{B}} \Phi (y)} + \left(g_{\mu \rho} - \frac{\partial^x_\mu \partial^x_\rho}{\partial_x^2}\right) \braket{g (\mathscr{A}^\rho \times \bar{\mathscr{C}})^A (x) \delta_{\text{B}} \Phi (y)},
 \label{eq:A-WI-expectation}
\end{align}
where we have used $\left(g_{\mu \rho} - \frac{\partial^x_\mu \partial^x_\rho}{\partial_x^2}\right) \partial_x^\rho = 0$. Using  $\frac{\delta S}{\delta \mathscr{B}^A (x)} = \partial_\rho \mathscr{A}^{\rho A} (x) + \alpha \mathscr{B}^A (x)$ and the left derivatives $\frac{\delta S}{\delta \mathscr{C}^A (x)} = - i (\mathscr{D}^\rho \partial_\rho \bar{\mathscr{C}})^A (x)$, the first term of \eqref{eq:A-WI-expectation} excluding $\frac{\partial^x_\mu}{\partial_x^2}$ is rewritten into
\begin{align}
 &\partial^x_\rho \braket{(\mathscr{D}^\rho \bar{\mathscr{C}})^A (x) \delta_{\text{B}} \Phi (y)} \nonumber\\
 = &\int d \mu \ e^{i S} \partial_\rho (\mathscr{D}^\rho \bar{\mathscr{C}})^A (x) \delta_{\text{B}} \Phi (y) \nonumber\\
 = &\int d \mu \ e^{i S} [(\mathscr{D}^\rho \partial_\rho \bar{\mathscr{C}})^A (x) \delta_{\text{B}} \Phi (y) + (g \partial_\rho \mathscr{A}^\rho \times \bar{\mathscr{C}})^A (x) \delta_{\text{B}} \Phi (y)] \nonumber\\
 = &\int d \mu \ e^{i S} \left[i \frac{\delta S}{\delta \mathscr{C}^A (x)} \delta_{\text{B}} \Phi (y) + \left(g \left(\frac{\delta S}{\delta \mathscr{B}} - \alpha \mathscr{B}\right) \times \bar{\mathscr{C}}\right)^A (x) \delta_{\text{B}} \Phi (y)\right].
 \label{eq:longitudinal}
\end{align}
The first term in the right-hand side of \eqref{eq:longitudinal} is simplified by using the Schwinger-Dyson equation $\int d \mu \frac{\delta}{\delta \Phi^A (x)} (\cdots) = 0$
\begin{align}
 &\int d \mu \ e^{i S} i \frac{\delta S}{\delta \mathscr{C}^A (x)} \delta_{\text{B}} \Phi (y) \nonumber\\
 = &\int d \mu \frac{\delta}{\delta \mathscr{C}^A (x)} [e^{i S} \delta_{\text{B}} \Phi (y)] - \int d \mu \ e^{i S} \frac{\delta}{\delta \mathscr{C}^A (x)} \delta_{\text{B}} \Phi (y) \nonumber\\
 = &- \Braket{\frac{\delta}{\delta \mathscr{C}^A (x)} \delta_{\text{B}} \Phi (y)}.
 \label{eq:schwinger_C}
\end{align}

Similarly, the second term in the right-hand side of \eqref{eq:longitudinal} reads
\begin{align}
 &\int d \mu \ e^{i S} \left(g \frac{\delta S}{\delta \mathscr{B}} \times \bar{\mathscr{C}}\right)^A (x) \delta_{\text{B}} \Phi (y) \nonumber\\
 = &\int d \mu \ e^{i S} g f^{A C E} \frac{\delta S}{\delta \mathscr{B}^C (x)} \bar{\mathscr{C}}^E (x) \delta_{\text{B}} \Phi (y) \nonumber\\
 = &\int d \mu \frac{\delta}{\delta i \mathscr{B}^C (x)} [e^{i S} g f^{A C E} \bar{\mathscr{C}}^E (x) \delta_{\text{B}} \Phi (y)] + i \int d \mu \ e^{i S} g f^{A C E} \bar{\mathscr{C}}^E (x) \frac{\delta}{\delta \mathscr{B}^C (x)} \delta_{\text{B}} \Phi (y) \nonumber\\
 = &- i g \Braket{\left(\bar{\mathscr{C}} \times \frac{\delta}{\delta \mathscr{B} (x)} \delta_{\text{B}} \Phi (y)\right)^A},
 \label{eq:schwinger_B}
\end{align}
which is supplemented with the trivial result:
\begin{align}
 - \int d \mu \ e^{i S} g \alpha (\mathscr{B} \times \bar{\mathscr{C}})^A (x) \delta_{\text{B}} \Phi (y) = g \alpha \braket{\delta_{\text{B}} (\mathscr{B} \times \bar{\mathscr{C}})^A (x) \Phi (y)} = 0.
\end{align}
Thus, \eqref{eq:A-WI2} is reduced to
\begin{align}
  &\int d^D x \ e^{i p (x - y)} i \partial^x_\mu \braket{\mathscr{J}_\omega^\mu (x) \Phi (y)} \nonumber\\
  = &\braket{\delta_\omega \Phi (y)} + \int d^D x \ e^{i p (x - y)} \partial^\mu \omega^A (x) \Biggl[ \frac{\partial^x_\mu}{\partial_x^2} \Braket{\frac{\delta}{\delta \mathscr{C}^A (x)} \delta_{\text{B}} \Phi (y)}  \nonumber\\
  &+ i g \frac{\partial^x_\mu}{\partial_x^2} \Braket{\left(\bar{\mathscr{C}} \times \frac{\delta}{\delta \mathscr{B} (x)} \delta_{\text{B}} \Phi (y)\right)^A} - \left(g_{\mu \rho} - \frac{\partial^x_\mu \partial^x_\rho}{\partial_x^2}\right) \braket{g (\mathscr{A}^\rho \times \bar{\mathscr{C}})^A (x) \delta_{\text{B}} \Phi (y)}\Biggr].
  \label{eq:WI_Phi}
\end{align}
In particular, if we take the single gauge field sector $\Phi (y) = \mathscr{A}_\nu^B (y)$, \eqref{eq:WI_Phi} is reduced to
\begin{align}
 &i \int d^D x \ e^{i p (x - y)} \partial_\mu^x \braket{\mathscr{J}_\omega^\mu (x) \mathscr{A}_\nu^B (y)} \nonumber\\
 = &\partial^\nu \omega^B (y) + \int d^D x \ e^{i p (x - y)} \partial^\mu \omega^A (x) \Biggl[\frac{\partial^x_\mu \partial^x_\nu}{\partial_x^2} \delta^{A B} \delta^D (x - y) \nonumber\\
 &- \left(g_{\mu \rho} - \frac{\partial^x_\mu \partial^x_\rho}{\partial_x^2}\right) \braket{g (\mathscr{A}^\rho \times \bar{\mathscr{C}})^A (x) (\mathscr{D}_\nu \mathscr{C})^B (y)}\Biggr].
\end{align}
Therefore, we obtain the restoration condition in the Lorenz gauge \eqref{eq:lorenz_case_restoration0}.


%
%

%


\bibliographystyle{unsrt}

\end{document}